\documentstyle[pre,aps,epsf,multicol]{revtex}
\begin{document}

\title{Phase transition of a two dimensional binary spreading model}

\author {G. \'Odor$^1$, M.C. Marques$^2$ and M.A. Santos$^2$\\}
\address{$^1$ Research Institute for Technical Physics and
         Materials Science, P. O. Box 49, H-1525 Budapest, Hungary}
\address{$^2$ Departamento de F\'\i sica and Centro de F\'\i sica do Porto, 
Faculdade de Ci\^encias, Universidade do Porto\\
Rua do Campo Alegre, 687 -- 4169-007 Porto -- Portugal}
\maketitle

\begin{abstract}
We investigated the phase transition behavior of a binary spreading process
in two dimensions for different particle diffusion strengths ($D$). We found
that $N>2$ cluster mean-field approximations must be 
considered to get consistent singular behavior. The $N=3,4$ approximations 
result in a continuous phase transition belonging to a single universality
class along the $D\in (0,1)$ phase transition line. Large scale simulations 
of the particle density confirmed mean-field scaling 
behavior with logarithmic corrections. This is interpreted as numerical 
evidence supporting that the upper critical dimension in this model is 
$d_c=2$. 
The pair density scales in a similar way but with an additional logarithmic
factor to the order parameter. At the $D=0$ endpoint of the transition 
line we found DP criticality.
\end{abstract}

\begin{multicols}{2}
\section{Introduction}

The study of nonequilibrium phase transitions in systems with absorbing phases
is an active area of research in Statistical Physics with applications in 
various other fields such as chemistry, biology and social sciences 
 \cite{Dick-Mar}. The classification of the types of phase transitions 
found in these systems into universality classes is, nevertheless, still an 
incomplete task. 

The directed percolation (DP) universality class is the most common 
nonequilibrium universality class \cite{Jan81,Gras82}. Directed percolation 
was indeed found to describe the critical behavior of a wide range of systems, 
despite the differences in their microscopic dynamic rules. However, the 
presence of some conservation laws and/or symmetries in the dynamics has been 
found to lead to other universality classes \cite{Hin2000}. 

The pair contact process (PCP)\cite{PCP} is one of the models whose 
(steady state) critical properties belong to the DP universality class. 
If the model is generalized to include single particle diffusion 
(PCPD or annihilation/fission model \cite{Car,HT97}), a qualitatively 
distinct situation arises since states with only isolated particles are 
no longer frozen and the question was raised of whether this would 
modify the universality class. A field theoretical study of the 
annihilation/fission model was presented long ago \cite{HT97}. 
Unfortunately, it relies on a perturbative renormalization group
analysis which breaks down in spatial dimensions $d \le 2$ so that the 
active phase and the phase transition are inaccessible to this study. 
The upper critical dimension $d_c$ is $2$ for this bosonic theory, where 
multiple site occupancy is allowed, contrary to the usual lattice models
and Monte Carlo simulations. A fermionic field theory is not available but 
it is expected to lead to $d_c=1$ \cite{Uunp} -- therefore mean field 
predictions, with some logarithmic corrections, would be seen in $d=1$ 
if the latter is the correct theory.

Monte Carlo, Coherent Anomaly \cite{Odo00,Hayepcpd} and DMRG studies 
\cite{Car,MHUS} of the $1-d$ PCPD proved to be rather hard due to very long 
relaxation times and important corrections to scaling. 
Several hypothesis were put forward in order to classify its critical 
behavior: single type \cite{Car,MHUS} versus two regions of different 
behavior \cite{Odo00}, parity conserving (PC) class \cite{Car}, 
mean field behavior, diffusion-dependent exponents. Some related models 
were also studied \cite{HH,Odo01,ParkH,mspread,HayeDP-ARW,Park} with 
the aim of identifying the relevant features which determine the critical 
properties. The matter is not yet fully clarified, but it seems more likely 
that this system belongs to a distinct, not previously encountered, 
universality class. Park {\it et al} \cite{ParkH} have also pointed out that 
the {\it binary} character of the particle creation mechanism, rather than 
parity conservation, might be the crucial factor determining the type of 
critical behavior of PCPD. Higher dimensional studies of PCPD-like models are 
thus necessary in order to clarify their universal properties and thus 
contribute to a full understanding of the nonequilibrium
phase transitions puzzle.

In the present work we have studied a two-dimensional model where particle 
diffusion competes with binary creation and annihilation of pairs of particles.
The model is described in the following section. Cluster mean field studies 
are presented in section III  and Monte Carlo simulations are discussed in 
section IV. Finally we summarize and discuss our results.

\section{The model}

The sites of a square lattice of side $L$ are either occupied 
by a particle ($1$) or empty ($0$). The following 
dynamic rules are then performed sequentially. 
A particle is selected at random and i) with probability $D$ 
is moved to a (randomly chosen) empty neighbor site; with the 
complementary probability, and provided the particle has at least 
one occupied neighbor, then ii) the two particles annihilate  
with probability $p$ or iii) with probability $1-p$ two particles 
are added at vacant neighbors of the initial particle. The selection 
of neighbors is always done with equal probabilities; the updating 
is aborted and another particle is selected, 
if the chosen sites are not empty/occupied as required by the process. 
In reaction-diffusion language one has
\begin{equation}
A\emptyset \stackrel {D}{\leftrightarrow}  \emptyset A , \ \ \ 
2A \stackrel {p(1-D)}{\longrightarrow} \emptyset , \ \ \
2A \stackrel {(1-p)(1-D)}{\longrightarrow}  4A
\end{equation}
It is clear that, in the absence of particle diffusion ($D=0$), 
only sites which belong to a pair of occupied neighbors are active. 
In terms of pairs -- taken as redefined 'particles' -- we have a {\it unary} 
process where 'particles' are destroyed at rate $p$ or give birth to 
new 'particles' at rate $1-p$; the number of offsprings is greater or 
equal to two -- because new pairs may also be formed with next nearest
neighbor particles -- and parity of the number of 'particles' is 
not conserved. 
This is similar to the PCP and one expects to see a phase transition, 
in the DP universality class, between an active phase with a finite 
density of pairs (at low $p$) and an absorbing phase without pairs 
(for $p>p_c$).

When particle diffusion is included, one has a qualitatively 
different situation, since configurations with only lonely particles 
are no longer absorbing -- the only absorbing states are the 
empty lattice and the configurations with a single particle. 
There is parity conservation in terms of particles and the creation 
and annihilation mechanisms are {\it binary}. The nature of the phase 
transition, expected to occur at some value $p_c(D)$,  is investigated below.  
 
\section{Cluster mean-field calculations} \label{sectclus}

We performed $N$-cluster mean-field calculations \cite{gut87,dic88} for 
this model. Since the details of the dynamics will not influence the values of 
the mean field critical indices, we have considered a simpler one dimensional 
version of the model where the creation takes place at the nearest and 
next-nearest sites to one side of the mother particle. 

At the site ($N=1$) level, the evolution of the particle density $\rho$
(denoted by $n$ in \cite{Car}) can be expressed as
\begin{equation}
\frac{\partial \rho}{\partial t}  =  -2 p \rho^2 +2(1-p) \rho^2 (1-\rho)^2
\end{equation}
which has a stationary solution
\begin{equation}
\rho(\infty) = \frac{p-1 + \sqrt{p-p^2}} {p-1}
\end{equation}
with $p_c=1/2$. The pair density $\rho_2$ ($c$ in \cite{Car} notation)  
is just the square of  $\rho$ at this level.
For $p<\sim p_c$ the densities behave as
\begin{eqnarray}
\rho(\infty)\propto (p_c-p)^{\beta} \\
\rho_2(\infty)\propto (p_c-p)^{\beta_2}
\end{eqnarray}
with $\beta=1$ and $\beta_2=2$ leading order singularities.
At the critical point 
\begin{equation}
\frac {\partial \rho(p=1/2)}{\partial t} = 2 (\rho/2-1) \rho^3
\end{equation}
which implies the leading order scaling is
\begin{equation}
\rho(t) \propto t^{-\alpha} \ , \ \ \ \rho_2(t) \propto t^{-\alpha_2} \ \ \ ,
\end{equation}
with $\alpha=1/2$ and $\alpha_2=1$, while in the absorbing phase 
\begin{equation}
\rho(t) \propto t^{-1} \ , \ \ \ \rho_2(t) \propto t^{-2}  \ \ \ .
\end{equation}
All these exponents coincide with those found for the PCPD model 
\cite{Car} at the same level of approximation.

In the pair ($N=2$) approximation, the density of '1' $\rho$ and the '11' 
pair density $\rho_2$ are independent quantities.
One can easily check that the evolution of particles can be expressed as
\begin{eqnarray}
\frac{\partial \rho}{\partial t}  & = & -2 p (1-D) \rho_2 + \nonumber \\ 
 & + & 2 (1-D) (1-p) \rho_2 (\rho-\rho_2) \frac{1-2\rho+\rho_2}{\rho(1-\rho)}
\end{eqnarray}
while the evolution of pairs
\begin{eqnarray}
\frac{\partial \rho_2}{\partial t} & = & -p (1-D) \rho_2 \frac{2 \rho_2+\rho}
{\rho} - 2 D (\rho-\rho_2) \frac{\rho_2-\rho^2}{\rho(1-\rho)} + \nonumber \\ 
& + &  (1-D) (1-p)  \rho_2 (\rho- \rho_2) (1-2 \rho+ \rho_2) 
\frac{2-\rho-\rho_2}{\rho(1-\rho)^2 }
\end{eqnarray}
Owing to the nonlinearities, we could not solve these equations analytically 
and had to look for numerical solutions. The critical indices thus obtained 
at different diffusion rates $D$ are shown in Table \ref{tab23}. 
As we can see, there are two distinct regions. For $D>\sim 0.2$ $p_c$ 
is constant and $\beta_2=2$, while for $D < \sim 0.2$  $p_c$ varies with 
$D$ and $\beta_2=1$. All these results are in complete agreement with 
those of the PCPD model in the pair approximation.

In the $N=3$ level approximation the situation changes, as we can see
in Table \ref{tab23}: the two distinct regions for $D>0$ disappear 
and $\beta_2=2$ everywhere as found in the site approximation. 
At $D=0$, however, the particle density does not vanish at the transition 
but goes to $\rho(p_c)=0.2931$. This means that the $N=3$ level 
approximation is already capable to describe the absorbing state 
that contains frozen, isolated particles. 
For $p< \sim p_c$, $\rho(p_c)-\rho \propto (p_c-p)^{\beta}$ 
with $\beta=1$, the same critical exponent as the order parameter 
(the pair density) therefore we redefine eq.(4) now. 
These results are also in agreement with those of the PCP model 
\cite{porto,pcph,natocikk}.

\begin{table}
\begin{center}
\begin{tabular}{|c|c|c|c|c|c|c|c|c|c|}
$D$ & \multicolumn{3}{c|}{$N=2$} & \multicolumn{3}{c|}{$N=3$} & 
\multicolumn{3}{c|}{$N=4$}\\
    &  $p_c$  & $\beta $ & $\beta_2$ & $p_c$ & $\beta $ & $\beta_2$ & $p_c$ & 
$\beta $ & $\beta_2$ \\
\hline
0.75      &  0.5   & 1  & 2 &  0.4597 & 1  & 2 & 0.4146 & 1 & 2 \\
0.5       &  0.5   & 1  & 2 &  0.4    & 1  & 2 & 0.3456 & 1 & 2 \\
0.25      &  0.5   & 1  & 2 &  0.3333 & 1  & 2 & 0.2973 & 1 & 2 \\
0.1       &  0.4074& 1  & 1 &  0.2975 & 1  & 2 & 0.2771 & 1 & 2 \\
0.01      &  0.3401& 1  & 1 &  0.2782 & 1  & 2 & 0.2759 & 1 & 2 \\
0.00      &  0.3333& 1  & 1 &  0.1464 & 1  & 1 & 0.1711 & 1 & 1 \\
\end{tabular}
\end{center}
\caption{Summary of $N=2,3,4$ approximation results\label{tab23}}
\end{table}

\begin{figure}
\epsfxsize=70mm
\epsffile{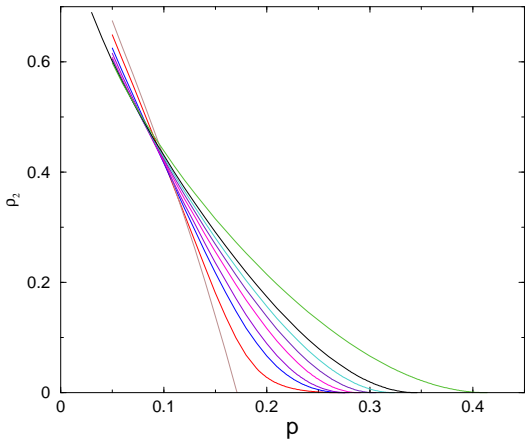}
\vspace{2mm}
\caption{$N=4$ cluster mean-field results for $\rho_2$. The curves correspond 
to $D=0,0.01,0.05$,$0.1$,$0.2$,$0.3$,$0.4$,$0.5$,$0.75$ (left to right). 
The same kind of convex curvature corresponding to $\beta_2=2$ can be 
observed for $D>0$, while it is different for $D=0$ (dashed line) 
corresponding to $\beta=1$.}
\label{afpc4}
\end{figure}

This kind of singular mean-field behavior persists for $N=4$ 
(Table \ref{tab23}, Fig.\ref{afpc4}) and can be found in the $N=3,4$ 
level approximations of the PCPD model as well \cite{pcpdnew}. 
These results suggest that the $N=2$ approximation is an odd one.
Recent investigations in similar PCP-like models \cite{porto,pcph} have also 
shown discrepancies in the singular behavior of the low-level cluster
mean-field approximations. One can conclude that in these models at least 
$N>2$ cluster approximations are necessary to find a correct 
mean-field behavior. It is probably just a coincidence that the $N=1$ 
calculation produced the same results. We shall thereafter ignore the $N=2$ 
results and refer to the $N=1,3$ scenario as the mean field prediction.

\section{Simulation results}

The simulations were started on small lattice sizes ($L=100,200$) to locate
the phase transition point roughly at $D=0,0.05,0.2,0.5,0.8$. The
particle density decay $\rho^L(t)$ was measured up to $t_{max} = 60000$ MCS
in systems started from fully occupied lattices and possessing periodic 
boundary conditions.
Throughout the whole paper $t$ is measured in units of Monte-Carlo sweeps 
(MCS). For $D=0.05$ we have not done so detailed analysis as for other
diffusion rates but only checked that the results are in agreement with 
the conclusions derived from the $D=0.2,0.5,0.8$ data.

Then we continued our survey on larger lattices: $L=400,500,1000,2000$ and 
determined $p_c$ at each size by analyzing the local slopes of $\rho(t)$
\begin{equation}
\alpha_{eff}(t) = {- \ln \left[ \rho(t) / \rho(t/m) \right] 
\over \ln(m)} \label{slopes}
\end{equation}
(we used $m=8$). In the $t\to\infty$ limit the critical curve 
goes to exponent $\alpha$ by a straight line, while sub(super)-critical 
curves veer down(up) respectively. The $p_c(L)$ estimates exhibit an 
increase with $L$ hence at the true critical point the critical like
$\alpha_{eff}$ curves of a given $L$ are sub-critical. This excludes the
possibility of finite size scaling study at $p_c$.

\subsection{Dynamical scaling for $D>0$} \label{dynsect}

\begin{figure}
\epsfxsize=80mm
\epsffile{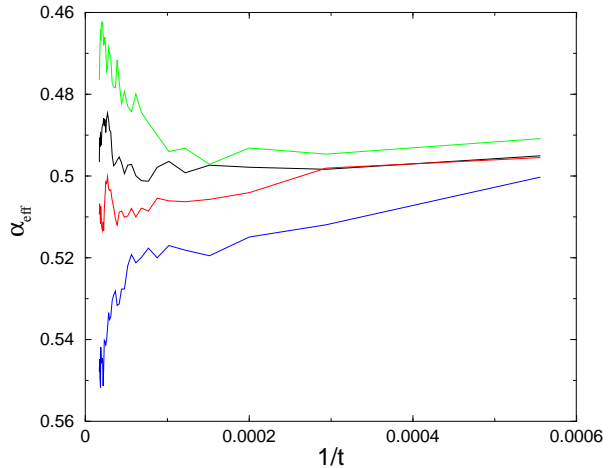}
\vspace{2mm}
\caption{Local slopes of the particle density decay at $D=0.5$ and
$L=2000$. Different curves correspond to $p=0.4392$, $0.43916$, $0.43913$, 
$0.4391$ (from bottom to top).}
\label{mariapl}
\end{figure}

\begin{figure}
\epsfxsize=80mm
\epsffile{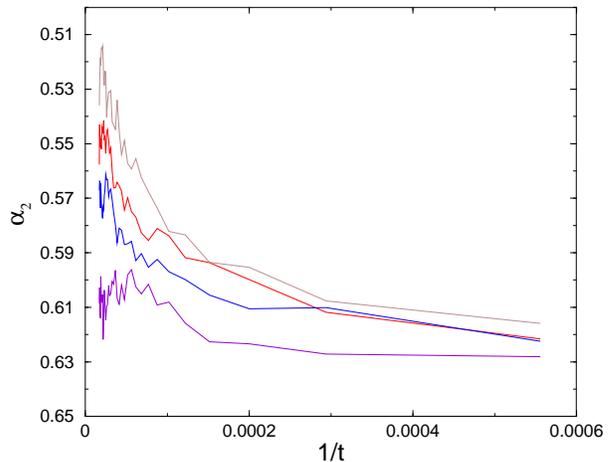}
\vspace{2mm}
\caption{The same as Fig.\ref{mariapl}. for $\rho_2(t)$.}
\label{mariapl2}
\end{figure}

\begin{figure}
\epsfxsize=80mm
\epsffile{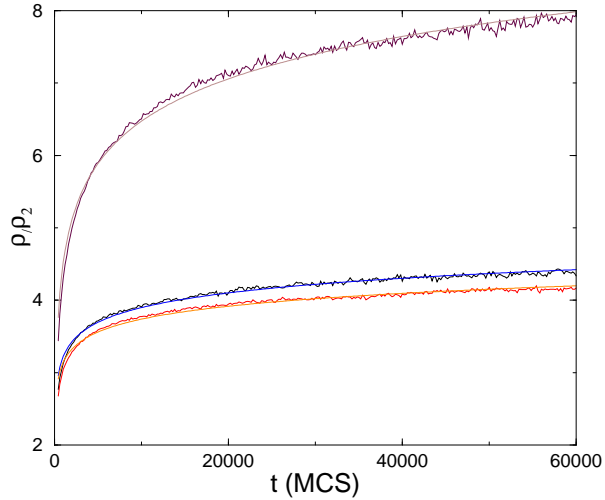}
\vspace{2mm}
\caption{$\rho(t)/\rho_2(t)$ and logarithmic fit for critical curves 
determined by $\rho(t)$ analysis. The top curves correspond to 
$D=0.8$, $p=0.475$, the middle ones to $D=0.2$, $p=0.41235$,
while the bottom ones correspond to $D=0,5$, $p=0.43916$. 
Note that the ratio is  smallest at $D=0.5$.}
\label{r1r2}
\end{figure}

For the largest system size ($L=2000$) at $D=0.5$ diffusion rate the local 
slopes analysis results in a $p_c=0.43915(1)$ and the corresponding 
$\alpha = 0.50(2)$ decay exponent agrees with the mean-field value 
(see Fig.\ref{mariapl}).
We also measured the pair density $\rho_2(t)$; applying a local slope analysis 
similar to eq.(\ref{slopes}) suggests (Fig.\ref{mariapl2}) the lack 
of phase transition of this quantity at $p=0.43915$.
Instead the curves veer up which may lead to different $p_c$ and 
$\alpha_2$ estimates. Such strange behavior has already been observed 
in PCPD model simulations \cite{Hunp,Gunp}.
An explanation for this discrepancy was pointed out by Grassberger in the 
case of the PCPD model \cite{Gunp}. Random walks in two dimensions are just 
barely recurrent and single particles can diffuse very long before they 
encounter other particles. Therefore it is natural to expect that 
$\rho(t)/\rho_2(t) \sim \ln(t)$ and Fig.\ref{r1r2} shows this really 
happens at $p_c$.
Therefore at $D=0.5$ we can conclude that $\alpha_2 \simeq \alpha \simeq 0.5$
taking into account logarithmic corrections. This however contradicts
the mean-field approximation value $\alpha_2=1$.

\begin{figure}
\epsfxsize=80mm
\epsffile{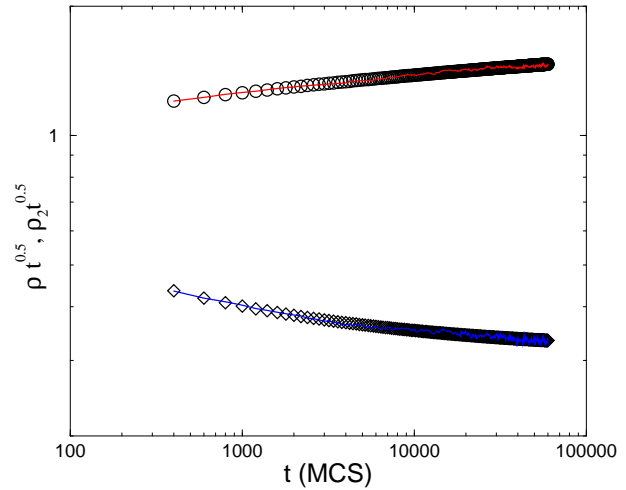}
\vspace{2mm}
\caption{Logarithmic fit (circles) for $\rho(t)$ (upper curve)
and with the form (\ref{ln2}) (diamonds) for $\rho_2(t)$ (lower curve) 
at $D=0.2$ and $p=0.41235$.}
\label{mariapl12}
\end{figure}

Similar local slopes analysis for $D=0.2$ and $D=0.8$ seem to imply
$\alpha=0.46(2)$ and $\alpha=0.57(2)$ respectively. First this raises the
idea that the exponents would change continuously with $D$ as it was 
observed in some one dimensional PCPD simulations \cite{Odo00,Park}.
Nevertheless the deviations from $0.5$ are small hence we tried to fit our 
data including logarithmic corrections. Logarithmic corrections may really 
arise if $d_c=2$ as predicted by bosonic field theory \cite{HT97}.
The precise form of these corrections is however not known for the present 
case so we have tried several functional dependences and found that  
\begin{equation}
((a+b\ln(t))/t)^{\alpha}
\label{lnform}
\end{equation}
is a good choice. As Fig.\ref{mariapl12} shows for $D=0.2$ this really works 
with $\alpha=0.507$ exponent.

Similarly for $D=0.8$ the same logarithmic formula fitting resulted in
$\alpha=0.497$. The coefficient of the logarithmic correction term is negative
($b=-0.2776$), while it is positive for $D=0.2$ ($b=0.468$). 
These results
suggest that logarithmic corrections to scaling should work for all cases we 
investigated but at $D=0.5$ they are very small and change sign.
Indeed applying the same formula for the $D=0.5$, $p=0.43913$ data we 
obtained $\alpha=0.496$ with $b=0.00027$ and $a=1.552$.
As we found logarithmic corrections to the particle density decay and a
logarithmic relation between $\rho_2(t)$ and $\rho(t)$ we may expect even 
stronger logarithmic corrections to the $\rho_2(t)$ data. 
Trying different forms for $D=0.2,0.5,0.8$ we found that taking into account 
$\ln^2(t)$ correction terms is really necessary and the best choice is
\begin{equation}
((a + b \ln(t)+ c\ln^2(t)) t)^{-\alpha} \ \ \ .
\label{ln2}
\end{equation}
This resulted in $\alpha_2=0.5007$ for $D=0.2$ (see Fig.\ref{mariapl12});
$\alpha=0.501$ for $D=0.5$ and $\alpha=0.484$ for $D=0.8$.
All these results imply that $\alpha=\alpha_2=0.5$ independently from the
diffusion rate $D$. For $\rho(t)$ this agrees with the mean-field
approximations and we don't see a change of universality by varying $D$
inferred from the $N=2$ approximation. The critical behavior
of $\rho_2(t)$ however differs from the $\alpha_2^{MF}=1$ prediction.

\subsection{Static behavior for $D>0$}

\begin{figure}
\epsfxsize=80mm
\epsffile{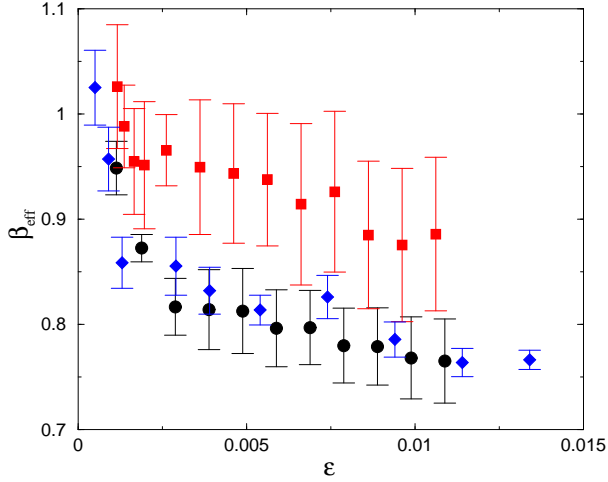}
\vspace{4mm}
\caption{Effective order parameter exponent results. Circles correspond
to $D=0.2$, diamonds to $D=0.5$, squares to $D=0.8$ data.}
\label{beta}
\end{figure}

\begin{figure}
\epsfxsize=80mm
\epsffile{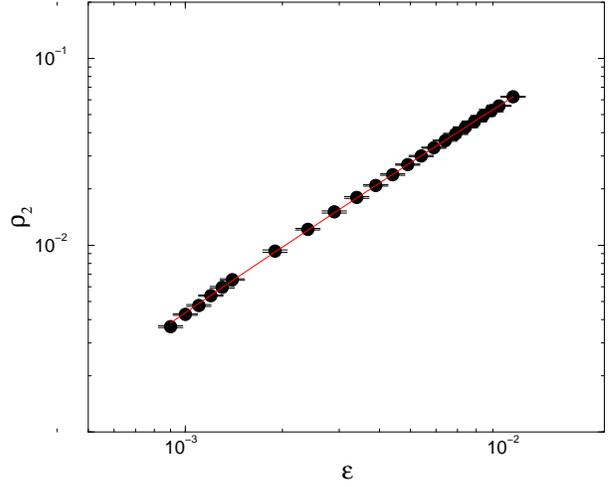}
\vspace{4mm}
\caption{Logarithmic fitting to $\rho_2(\infty)$ at $D=0.5$ using the form
eq.(\ref{logform}). The coefficients are $a=0.112$, $b=0.01$ and
$\beta=0.96(5)$.}
\label{betal}
\end{figure}

The $p_c$ estimates for different sizes were used to extrapolate to the
true critical value. Simple linear fitting as a function of $1/L$ resulted 
in the values given in Table \ref{tab}. For determining steady state exponents 
the densities $\rho^L(t,p,D)$ and $\rho^L_2(t,p,D)$ were followed in the 
active phase until level-off values were found to be stable. 
Averaging was done in the level-off region for $100-1000$ surviving samples 
-- those with more than one particle \cite{comment}.
Again at each $p$ and $D$ we extrapolated as a function of $1/L$ to the 
$\lim_{L\to\infty}\rho^L(\infty,p,D)$ values.
The local slope analysis of exponent $\beta$
\begin{equation}
\beta_{eff}(\epsilon_i,D) = \frac {\ln[\rho(\infty,\epsilon_i,D)] -
\ln[\rho(\infty,\epsilon_{i-1},D)]}{\ln(\epsilon_i)-\ln(\epsilon_{i-1})} \ \ ,
\end{equation}
(where $\epsilon=p_c-p$) shows that the order parameter $\rho$ exhibits 
a $\beta\simeq 1$ asymptotic scaling at $D=0.2,0.5,0.8$ (Fig.\ref{beta})
although a correction to scaling can be seen in all cases.
The $\beta\simeq 1$ agrees with the mean-field prediction.
Doing the same analysis for $\rho_2$ the local slopes seem to extrapolate
to $\beta\simeq 1.2$ for each $D$. This is very far from the mean-field value
$\beta_2^{MF}=2$ and we don't see any change by varying the
diffusion rate down to $D=0.05$.
We have investigated the possibility of different logarithmic corrections 
and found that the
\begin{equation}
\rho = (\epsilon / (a + b \ln(\epsilon))^{\beta}
\label{logform}
\end{equation}
form gives very good fitting with $\beta=0.96(5)$ for $D=0.5$ while
for $D=0.2$ and $D=0.8$ (similarly to the exponent $\alpha_2$ case) we need 
to take into account $\ln^2(\epsilon)$ correction terms to obtain similar good 
fitting (see Table \ref{tab} and Fig.\ref{betal}).
Therefore we concluded that, as in  $\alpha$ case, the steady state 
exponents are equal: $\beta=\beta_2$.
Note that we have checked that logarithmic corrections to scaling can also
be detected in $\rho$ data with $\beta\simeq 1$ exponent.

\subsection{Data collapse for $D>0$}

\begin{figure}
\epsfxsize=80mm
\epsffile{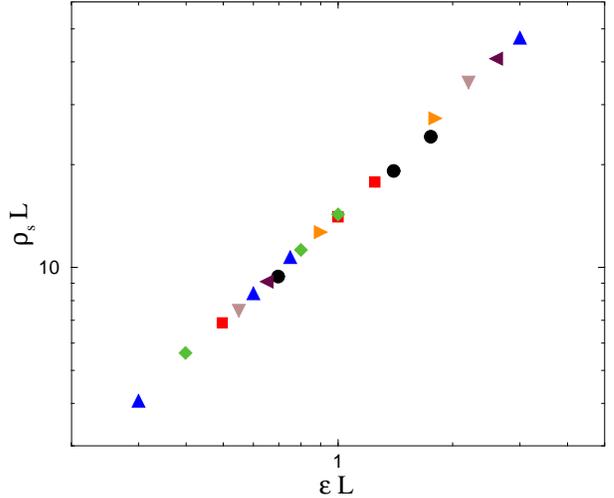}
\vspace{4mm}
\caption{Finite size data collapse according to scaling form (\ref{fssform})
for $L=200$,$400$,$500$,$1000$,$2000$. Different symbols denote data for 
$p=0.436$,$0.437$,$0.4375$,$0.438$,$0.4382$,$0.4384$,$0.4386$}
\label{fss_4395}
\end{figure}

To test further the possibility of mean-field critical behavior we performed 
finite size scaling on our $\rho^L(t,p,D)$ data assuming the mean-field 
exponents \cite{HT97} $\beta=1$, $\nu_{\perp}=1$ and the scaling form
\begin{equation}
\rho^L(\infty,\epsilon,D) \propto L^{-\beta/\nu_{\perp}} 
f(\epsilon L^{1/\nu_{\perp}}) \ \ \ .
\label{fssform}
\end{equation}
As Fig.\ref{fss_4395} shows, a good data collapse was obtained for
$p_c=0.4395(1)$ at $D=0.5$.
Similarly the scaling form
\begin{equation}
\rho^L(t,\epsilon,D) \propto t^{-\beta/\nu_{||}} 
g(t \epsilon^{\nu_{||}}) 
\label{scform}
\end{equation}
($\alpha = \beta/\nu_{||}$) can be checked near $p_c$, assuming the 
mean-field values \cite{HT97} $\beta=1$ and $\nu_{||}=2$. 
For the largest size ($L=2000$) at $D=0.5$ the best collapse of curves 
corresponding to $p=0.438$,$0.4382$,$0.4384$,$0.4386$,$0.4388$ was obtained
for $p_c=0.4394(1)$. This agrees well with previous $p_c$ estimates within
margin of numerical accuracy.
\begin{figure}
\epsfxsize=80mm
\epsffile{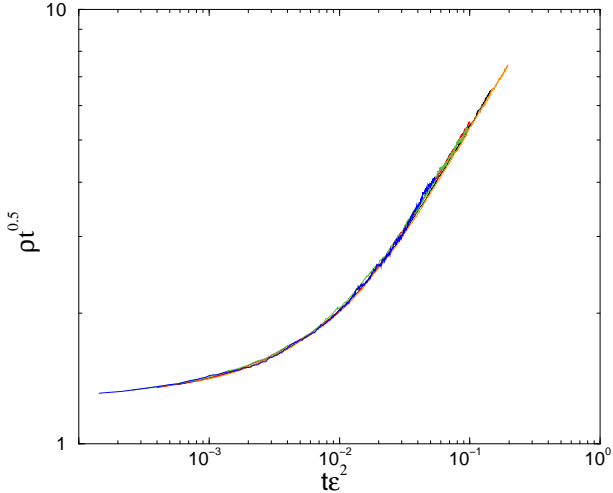}
\vspace{4mm}
\caption{Data collapse according to scaling form (\ref{scform}) at
$D=0.5$. Different curves correspond to data at 
$p=0.438$,$0.4382$,$0.4384$,$0.4386$,$0.4388$}
\label{col94}
\end{figure}

\subsection{The $D=0$ case}

As explained above, we expect that this model exhibits 2+1 dimensional DP 
universality because for the pair density the conditions of the DP 
hypothesis \cite{Jan81,Gras82} are satisfied. Indeed at $p_c=0.3709(1)$ we 
found that the decay exponent of pairs is $\alpha_2=0.45(1)$ and the steady 
state density approaches zero with the scaling exponent $\beta_2=0.582(1)$ in 
agreement with the estimates for this class $\alpha=0.4505(10)$ and 
$\beta=0.583(14)$ \cite{DP2}. At the critical point the density of isolated 
particles takes a nonzero value, usually called the {\it natural} density, 
$\rho(p_c) \simeq 0.135$. In \cite{natocikk} we showed that in case of the
PCP and an other 1d model exhibiting infinitely many absorbing states the 
nonorder field follows the scaling of the order parameter field. Here we 
found that the total density shows a singular behavior
\begin{equation}
 \rho(p)-\rho(p_c) \propto (p_c-p)^{\beta}
\end{equation}
with the redefined exponent $\beta=0.60(2)$ agreeing with that of the DP 
class within margin of numerical accuracy.

\subsection{Scaling in the inactive phase}

\begin{figure}
\epsfxsize=80mm
\epsffile{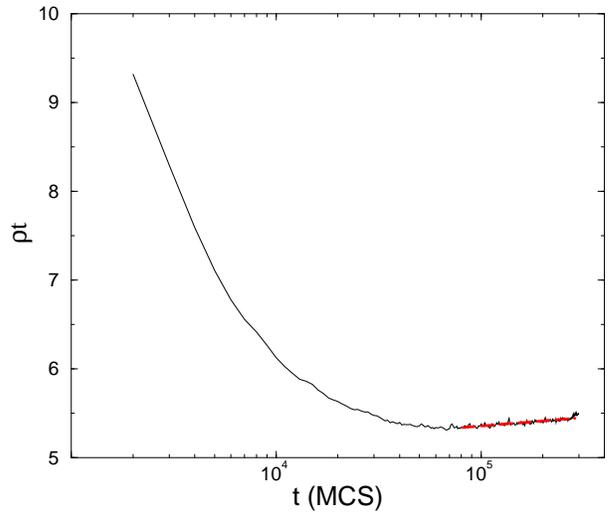}
\vspace{4mm}
\caption{Density decay in the inactive phase. For large times we found
$\rho(t) = (4.46 + 0.078(2) \ln t) /t$ behavior (dashed line)
by fitting data.}
\label{inactive}
\end{figure}

According to the bosonic field theory \cite{HT97} in the inactive phase the 
$A+A\to\emptyset$ reaction governs the particle density decay.
This process was solved exactly by Lee \cite{Lee} who predicted  the 
following late time scaling behavior in $d=2$
\begin{equation}
\rho(t) = \frac{1}{8\pi D} \ln t/t + O(1/t) 
\label{dec}
\end{equation}
We measured $\rho(t)$ at $p=0.45$ and $D=0.5$ in a $L=2000$ system 
up to $t_{max}=3\times 10^5$ MCS. As Figure \ref{inactive} shows,
for intermediate times the density decays faster than this power-law 
in agreement with results for PCPD \cite{Gunp} but later crosses over 
to the expected eq.(\ref{dec}) behavior with amplitude $0.078(2)$ and 
with a $4.46/t$ correction to scaling term.

Unlike what we found at the critical point (see section B), $\rho_2(t)$ 
decays faster than $\rho(t)$ in the absorbing phase.
The long-time behavior seems to be $\rho_2\propto t^{-2}$ which agrees with
the mean-field prediction.

\begin{table}
\begin{center}
\begin{tabular}{|l||r|r|r|}
           & $D=0.2$     & $D=0.5$    & $D=0.8$ \\
\hline
$p_c$      & 0.4124(1)   & 0.4394(1) & 0.4751(1)\\
$\alpha$   & 0.507(10)   & 0.496(6)  & 0.497(10)\\
$\alpha_2$ & 0.501(10)   & 0.501(5)  & 0.484(15)\\
$\beta$    & 1.07(10)    & 1.01(10)  & 1.07(10) \\
$\beta_2$  & 1.03(8)     & 0.96(5)   & 0.95(5) \\
\end{tabular}
\end{center}
\caption{Summary of simulation results at criticality\label{tab}}
\end{table} 

\section{Conclusions}

We have investigated the phase transition of a two dimensional binary spreading
model exhibiting parity conservation. In what concerns cluster mean-field
approaches, the results are similar to those of the PCPD model at 
the corresponding level of approximation \cite{Car,pcpdnew}. 
The $N=2$ results suggest two different universality classes depending on the 
diffusion strength. Higher ($N=3,4$) order cluster mean-field show a single 
universality class characterized by $\beta=1$ and $\alpha=1/2$. Comparing 
these with other recent results for PCP-like models and with the simulations, 
we believe that the $N=2$ case yields spurious results --- although two 
universality classes were apparently observed in a study of the one 
dimensional PCPD \cite{Odo00} --- and so $N>2$ cluster approximations 
are necessary to describe the mean-field singularity correctly. 
This is not surprising and was already found in similar models 
\cite{porto,pcph}. 
Note that in both the $N=3$ and $N=4$ approximations the $p_c$ seems to have
a discontinuity by approaching $D=0$. Similar discontinuity in the phase space
of the PCP model was recently reported as the result of an external particle
source \cite{pcph}. This behavior may be the subject of further studies.

We performed extensive and detailed simulations along the phase transition line
and found a single universality class with the order parameter exponents 
$\beta=1$ and $\alpha=0.5$ for all $D>0$. Logarithmic corrections to scaling 
were detected that are weakest at $D=0.5$. In the lack of a theoretical 
prediction, we have selected the best logarithmic fitting forms taking 
into account up to $O(\ln^2)$ terms, but we cannot rule out the possibility of 
other logarithmic correction forms. Scaling function analysis confirmed the
$\nu_{\perp}=2$ and $\nu_{||}=1$ mean-field values. This seems to indicate 
that the critical dimension is $d_c=2$ as predicted by the bosonic field 
theory. In the inactive region, the decay of particle density at large 
times was found to agree with an exact prediction \cite{Lee}

The pair density $\rho_2$ for $p\le p_c$ (where the bosonic field theory
breaks down) was shown to exhibit the same singular behavior as the order 
parameter, apart from a logarithmic ratio. Simulation results of the PCPD 
model \cite{Gunp,Hayepcpd} found indications for similar behavior. 
The reason why the mean-field approximation fails to describe the singular 
behavior of $\rho_2$ is yet not clear to us but in the two-component 
description of the model it indicates strong coupling between pairs and 
particles (similarly to other models \cite{PCP,PCPe}). 
In the inactive region, however, $\rho_2$ and $\rho$ scale differently. 
At the $D=0$ endpoint of the transition line we found 2+1 dimensional DP 
critical behavior of $\rho_2$ with infinitely many frozen absorbing states
similarly to the PCP model.

We have found identical predictions for the present and the PCPD model within 
mean-field, which seem to be confirmed by our simulations -- and also by 
preliminary simulations for the $2-d$ PCPD \cite{Hunp,Gunp}. One thus 
concludes that it is very likely that parity conservation is irrelevant for 
this transition, as in the one-dimensional case \cite{ParkH} and in certain 
models with exclusion \cite{excl}. Further renormalization group studies of 
these systems are necessary for a proper justification of these results.

\noindent
{\bf Acknowledgements:}\\
We thank P. Grassberger, H. Hinrichsen and U. C. T\"auber for 
communicating their unpublished results and M. Henkel for his comments.
Support from Hungarian research funds OTKA (Grant No. T-25286), Bolyai 
(Grant No. BO/00142/99) and IKTA (Project No. 00111/2000) and 
from project POCTI/1999/Fis/33141 (FCT - Portugal) is acknowledged.  
The simulations were performed on the parallel cluster of SZTAKI and on the 
supercomputer of NIIF Hungary.

\end{multicols}
\end{document}